\documentclass[twocolumn,pre,showpacs]{revtex4} \usepackage{graphicx}
\usepackage{amsmath} \usepackage{amsfonts} 

\begin{document}
\title{Simplest nonequilibrium phase transition into an absorbing state}

\author{A. C. Barato$^1$, Juan A. Bonachela$^2$, C. E. Fiore$^3$, H.
  Hinrichsen$^1$, Miguel A. Mu{\~n}oz$^2$}

\affiliation{$^1$Universit\"at W\"urzburg, Fakult\"at f\"ur Physik und
  Astronomie, 97074 W\"urzburg, Germany,}

\affiliation{$^2$Departamento de Electromagnetismo y F{\'i}sica
         de la Materia and Instituto Carlos I de F{\'i}sica Te{\'o}rica y
         Computacional Facultad de Ciencias, Universidad de
         Granada, 18071 Granada, Spain}

\affiliation{$^3$Departamento de F\'\i sica, Universidade Federal do Paran\'a,
  Caixa Postal 19044, 81431 Curitiba, Paran\'a, Brazil }

\begin{abstract}
  We study in further detail particle models displaying a boundary-induced
  absorbing state phase transition [Phys. Rev.  E.  {\bf 65}, 046104 (2002)
    and Phys. Rev. Lett. {\bf 100}, 165701 (2008)] . These are one-dimensional
  systems consisting of a single site (the boundary) where creation and
  annihilation of particles occur, and a bulk where particles move diffusively.
  We study different versions of these models, and confirm that, except for
  one exactly solvable bosonic variant exhibiting a discontinuous transition
  and trivial exponents, all the others display non-trivial behavior, with
  critical exponents differing from their mean-field values, representing a
  universality class.  Finally, the relation of these systems with a
  $(0+1)$-dimensional non-Markovian process is discussed.
\end{abstract}
\pacs{64.60.Ht, 68.35.Rh, 64.70.-p}

\maketitle
\parskip 1mm

\section{Introduction}
\label{intro}

Phase transitions occurring in the bulk, but driven by specific conditions at
its boundaries, are called {\it boundary-induced phase
  transitions}~\cite{HenkelSchuetz94}.  Examples include diffusive
transport~\cite{Krug91,Schuetz00} and traffic flow~\cite{PopkovEtAl01}
models. A simple example for this is provided by the one-dimensional totally
asymmetric simple exclusion process \cite{Derrida93}, where particles enter
the system at the left boundary, jump to the right in the bulk, and exit at
the right boundary. Depending on the entering and exiting rate values, the
system exhibits qualitatively different phenomenologies (maximal current,
large current and low density, or small current and high density), with
straightforward applications to traffic flow problems.

In the present work, we are interested in boundary-induced phase transitions in
systems with absorbing states.  An absorbing state is a dynamical trap which
can be accessed but cannot be left~\cite{Hinrichsen00,Lubeck04,Odor04}.
Systems with absorbing phase transitions are controlled by a parameter,
depending on which the system either enters the absorbing state with certainty
or survives in a stationary fluctuating/active state. The most prominent
family of phase transitions into an absorbing state is the very robust direct
percolation (DP) universality class.  A recent breakthrough has been the
experimental observation of DP critical behavior for the first
time~\cite{TakeuchiEtAl07}.

A paradigmatic model in the DP universality class is the contact process (CP)
\cite{MarroDickman99}. It can be viewed as a simple model for the propagation
of a disease where sick individuals can infect healthy neighbors or become
healthy spontaneously. More precisely, in the CP in $d$ spatial dimensions, a
particle (infected individual) can be created at an ``empty'' site with a rate
$\lambda n/2d$, where $n$ is the number of nearest neighbors occupied by a
particle, and an occupied site can become empty at rate $1$.  The empty
configuration is an absorbing state.  For $\lambda$ larger than a certain
critical threshold, $\lambda_c$, the process is able to sustain (in an
infinite lattice) a non-vanishing density of particles, while for
$\lambda<\lambda_c$ the dynamics ends up, ineluctably, in the absorbing state.

As continuous phase transitions involve long-range correlations, boundary
effects may play an important role. In the context of absorbing phase
transitions, previous studies focused primarily on DP confined to
parabolas~\cite{KaiserTurban94,KaiserTurban95}, active
walls~\cite{HinrichsenKoduvely98}, as well as absorbing walls and
edges~\cite{Froedh98a,Froedh01a}. Although such boundaries influence the
dynamics deep into the bulk, the universality class of the bulk transition is
not inherently changed, rather it is extended by an additional independent
exponent describing the order parameter near the boundary.  Therefore, the
question arises whether it is possible to find boundary-induced absorbing
phase transitions, absent in the corresponding systems without boundaries,
constituting independent universality classes.

In this paper we present a detailed discussion of two slightly different
models introduced in Ref.~\cite{BaratoHinrichsen08} and
~\cite{DeloubriereWijland02}, respectively.  Both of them exhibit a
boundary-induced nonequilibrium phase transition into an absorbing
state. These are one-dimensional particle systems consisting of a single site
(and, at most, its nearest neighbor), where creation and annihilation of
particles occur, and a bulk, where particles move diffusively. While in the
first reference \cite{BaratoHinrichsen08}, the dynamics at the boundary is a
contact process, in the second one~\cite{DeloubriereWijland02} particles at
the origin annihilate only pairwise ($A+A \rightarrow 0$).  We study different
versions of these models to compare them and scrutinize the relevance of
relaxing the fermionic constraint (i.e. occupation number not restricted to
$0$ or $1$) both at the bulk and at the boundary.  Our study includes mean
field approximations, numerical analysis, some field theoretical arguments, as
well as the relation with a (0+1)-dimensional non-Markovian
model~\cite{BaratoHinrichsen09}.

The paper is organized as follows: In the next section we define the first
model and present numerical results. In Sec.~\ref{mean_field} we discuss
various types of mean field approximations and show that this model has indeed
a non-trivial behavior. Sec.~\ref{related_models} is concerned with two
bosonic versions of the first model and the study of models with pair
annihilation at the boundary. One of the bosonic versions is solved exactly
and it is shown to have trivial critical behavior. Instead, the other bosonic
version and models with pair annihilation are shown to share the same critical
behavior as the first model. In Sec.~\ref{non-Markovian} we discuss the
relation with a (0+1)-dimensional non-Markovian
model~\cite{BaratoHinrichsen09} and, finally, we present our main conclusions.

\section{Basic model definition and simulations}
\label{simulations}
\subsection{Definition of the model}
\label{definition}
\begin{figure}
\includegraphics[width=70mm]{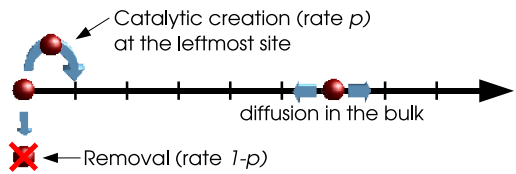}
\vspace{-2mm}
\caption{(Color online) Model on a semi-infinite lattice with symmetric
  diffusion in the bulk and special dynamical rules at the left boundary (see
  main text).
 \label{model}}
\end{figure}
The model presented in \cite{BaratoHinrichsen08} is defined on a
one-dimensional semi-infinite discrete lattice where each site is either
occupied by a particle ($s_i=1$) or empty ($s_i=0$). All lattice sites have
two neighbors, except for the boundary ($i=0$) with a single one. The dynamics
is a combination of an unbiased random walk in the bulk and a contact
process-like dynamics at the left boundary. It is implemented as follows:
\begin{enumerate}
\item[(a)] A particle is randomly selected.
\item[(b)] If it is located at the leftmost site, it generates another particle
  at site $1$ with probability $p$, provided that it is empty ($s_1=0$), or it
  dies ($s_0=0$) with probability $1-p$.
\item[(c)] Particles in the bulk perform a symmetric exclusion process,
  moving to any of their two neighbors with equal probability, provided that
  the destination site is empty (otherwise nothing happens).
\end{enumerate}
Starting with a single particle at the leftmost site in an otherwise absorbing
(i.e. empty) configuration, the process evolves as follows: the initial
particle at site $0$ either dies or generates another particle at the
neighboring site $1$. This last performs a random walk in the bulk until,
eventually, it returns to the origin to create another offspring or
disappear. A critical point, located at $p_c= 0.74435(15)$ has been reported
to separate the absorbing phase, in which the total number of particles
vanishes, from another with indefinitely sustained activity
\cite{BaratoHinrichsen08}.
\subsection{Order parameters}
\label{order_parameters}
\begin{figure}
\includegraphics[width=87mm]{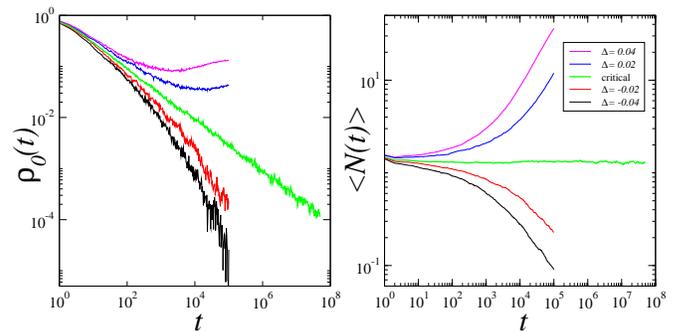}
\vspace{-2mm}
\caption{(Color online) Density of particles at the leftmost site
  $\rho_0$ (left) and the average total number of particles $N$
  (right) as functions of time for different values of $\Delta$,
  below, above, and at criticality. At the critical point,
  $\rho_0(t)$ decays as $t^{-1/2}$ while and $\langle N(t)\rangle$ is
  essentially constant.
  \label{fig:decay}}
\end{figure}
A possible order parameter for this model is the average density of
particles at the leftmost site:
\begin{equation}
  \rho_0= \langle s_0 \rangle\,,
\end{equation}
where $\langle \rangle$ stands for ensemble averages. This quantity is plotted
in Fig.~\ref{fig:decay} as a function of $\Delta:=p-p_c$. At the critical
point, $\rho_0(t)$ decays algebraically in time, as:
\begin{equation} 
\label{rhodecay}
\rho_0(t)\;\sim\; t^{-\alpha}
\end{equation}
with an exponent $\alpha= 0.50(1)$, compatible with a rational value $\alpha=
1/2$.

Another possibility is to choose as an order parameter the average total
number of particles, $\langle N(t) \rangle$, which, as shown in
Fig.~\ref{fig:decay}, goes to zero for $p<p_c$ and increases steadily for
$p>p_c$ (actually, it is limited only by the system size).  At criticality,
$\langle N (t) \rangle$ is found to be constant in the large time limit.

In the usual scaling picture of absorbing phase transitions, the critical
exponent $\beta$ is related to the probability that a given site belongs to an
infinite cluster generated from a fully occupied lattice at $t=-\infty$. This
quantity tends to zero as the control parameter approaches the critical value
from above. Similarly, the exponent $\beta'$ is related to the probability
that a localized seed generates an infinite cluster extending to
$t=+\infty$. Therefore, in the supercritical phase $(\Delta >0)$, the averaged
activity of the site at the origin for $t\to\infty$ \textit{measured in seed
  simulations} averaging over all runs, scales as $\rho^s \sim
\Delta^{\beta+\beta'}$, where the superscript `s' stands for `stationary'.  At
criticality, this function is expected to decay as $\rho(t) \sim
t^{-(\beta+\beta')/\nu_\parallel}$, where $\nu_\parallel$ is the correlation
time exponent.  Moreover, in the DP class a special {\it time reversal
  symmetry} implies that $\beta=\beta'$ \cite{Hinrichsen00}.

As shown in~\cite{DeloubriereWijland02}, time reversal symmetry also holds in
the present type of models. This implies that, in supercritical seed
simulations, the density of active sites at the boundary is expected to
saturate as:
\begin{equation}
  \rho_0^s\sim \Delta^{2\beta}\,,
\end{equation}
while, at criticality:
\begin{equation}
  \rho_0(t) \sim t^{-2\beta/\nu_\parallel},
\end{equation}
implying that $\alpha$ in Eq.~(\ref{rhodecay}) is
\begin{equation}
\alpha=2\beta/\nu_\parallel\,.
\label{alpha}
\end{equation}
Assuming that $\alpha=1/2$, then $\beta/\nu_\parallel = 1/4$.
\begin{figure}
\includegraphics[width=87mm]{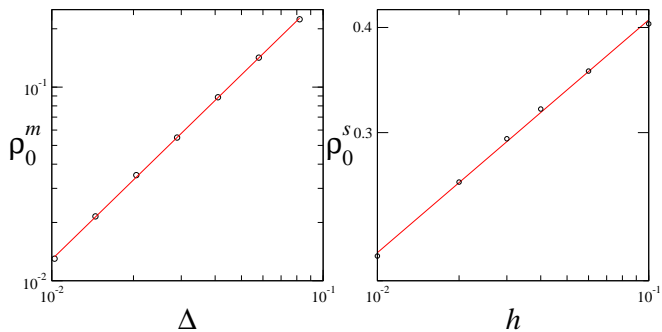}
\vspace{2mm}
\caption{(Color online) Left panel: Density of particles at the leftmost site,
  at the time when it reaches its minimum value, as a function of the distance
  from criticality $\Delta$. This gives the exponent $\beta= 0.68(5)$.  Right
  panel: The same quantity, at criticality and in the stationary state, as a
  function of the external field $h$, giving $\delta_h^{-1}= 0.29(5)$, compatible
  with the conjectured value $1/3$.
\label{fig:off}}
\end{figure}

\subsection{Stationary properties}
\label{stationary}
In numerical simulations in the active phase, it takes a very long time,
specially for small values of $\Delta$, to reach the steady state.  Moreover,
we observed the unusual fact that, for $\Delta>0$, the density $\rho_0$ goes
through a minimum before reaching the stationary state (see
Fig.~\ref{fig:decay} and also \cite{FES}, where similar non-monotonous curves
were reported). However, it turns out that the value $\rho_0^m$ at the minimum
and the saturation value $\rho_0^s$ differ by a constant factor, entailing
that both quantities scale in the same way, i.e.:
\begin{equation}
\rho_0^m\sim \Delta^{2\beta} \,.
\end{equation}
Note that this can be true only if the density $\rho_0(t)$ in seed simulations
obeys the scaling relation:
\begin{equation}
\rho_0(t) = \Delta^{2\beta}\, R(t\Delta^{\nu_\parallel})
\end{equation}
i.e. if it is possible to collapse the data by plotting
$\rho_0\Delta^{-2\beta}$ versus $t\Delta^{4\beta}$. Indeed, this will be shown
to be the case in Sec.~\ref{non-Markovian} for a $0$-dimensional non-Markovian
process argued to be in the same universality class.

Relying on this observation, one can determine the value of the exponent
$\beta$ by measuring the density $\rho_0^m$ at the minimum, which is reached
much earlier than the stationary state. In Fig.~\ref{fig:off} we plot $\rho_m$
as a function of $\Delta$, inferring $\beta= 0.68(5)$.

\subsection{External field}
\label{external}
In ordinary directed percolation, an external field, conjugate to the order
parameter, can be implemented by creating active sites at some constant rate
$h$, thereby destroying the absorbing nature of the empty configuration. At
criticality, the external field is known to drive a $d$+1-dimensional DP
process towards a stationary state with $\rho^s \sim h^{1/\delta_h}$ where
$\delta_h^{-1}=\beta/(\nu_\parallel+d\nu_\perp-\beta')$, and $\nu_\perp$ is the
correlation length critical exponent.

In the present model, the external field, conjugate to the order parameter
$\rho_0$, corresponds to spontaneous creation of activity at the leftmost site
at rate $h$. The above hyperscaling relation for $\delta_{h}$ is thus expected
to be fulfilled by taking $d=0$:
%
\begin{equation}
  \rho_0^s\sim h^{1/\delta_h}.
\end{equation}
with
\begin{equation}
  \delta_h^{-1}= \beta/(\nu_\parallel-\beta').
\end{equation}
From this expression, exploiting the fact that $\beta=\beta'$ and using
Eq.(\ref{alpha}) as well as the conjectured rational value $\alpha= 1/2$, a
prediction $\delta_h^{-1}=1/3$ is obtained. Our numerical estimate, $\delta_h^{-1}=
0.29(5)$ (see Fig.~\ref{fig:off}) is compatible with this result.

\subsection{Survival probability}
\label{survival}

\begin{figure}
\includegraphics[width=87mm]{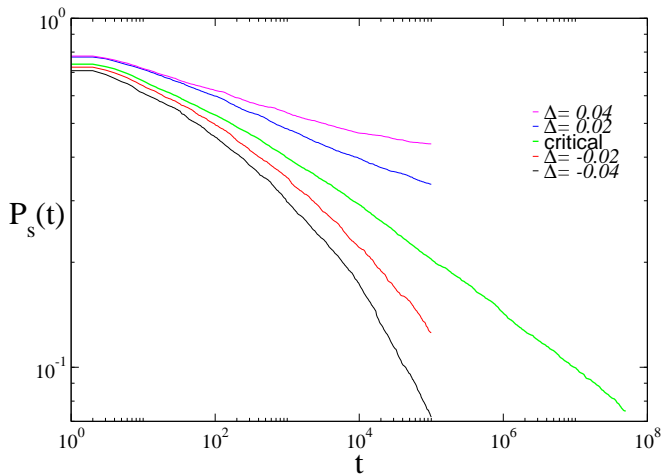}
\vspace{-2mm}
\caption{(Color online) Survival probability $P_s(t)$ as function of time
  below, above, and at criticality.  At the critical point, it decays with the
  exponent $\delta=0.15(2)$, different from $\beta/\nu_\parallel$, and in
  agreement with the conjectured value $1/6$.
\label{fig:survival}}
\end{figure}
The survival probability $P_s(t)$ is defined as the fraction of runs that,
starting with a single seed at the boundary, survive \textit{at least} until
time $t$. At criticality, this quantity is expected to decay algebraically:
\begin{equation}
P_s(t)\sim t^{-\delta},
\end{equation}
with the so-called survival exponent $\delta$, while in the super-critical
regime it saturates in the long time limit. Since $P_s(\infty)$ coincides with
the probability for a seed to generate an infinite cluster, the saturation
value of the survival probability as a function of the distance from
criticality gives the exponent $\beta'$. As in DP, one expects $P_s(t)$ to
decay in time with an exponent $\delta=\beta'/\nu_\parallel =1/4$. However, as
shown in Fig.~\ref{fig:survival}, one finds a much smaller exponent $\delta=
0.15(2)$. Therefore, the usual relation $\delta= \beta'/\nu_\parallel$ does
not hold. We also observed that it is not possible to collapse different
curves of $P_s(t)$ for different values of $\Delta$, i.e. the survival
probability seems to exhibit an anomalous type of scaling behavior. We expect
that off-critical simulations of the survival probability give the exponent
$\beta'$ but the simulation times needed to reach steady state are
prohibitively long.

An explanation for the value $\delta= 0.15(2)$, differing from
$\beta/\nu_\parallel$, is given in the following subsection.

\subsection{Time reversal symmetry}

\label{time_reversal}

In ordinary bond DP, the statistical weight of a configuration of percolating
paths does not depend on the direction of time. More specifically, the
probability to find an open path from at least one site at time $t=0$ to a
particular site at time $t$ coincides with the probability to find an open
path from a particular site at time $t=0$ to at least one site at time
$t$. This implies that, in bond DP, {\it i)} the density $\rho(t)$ in simulations with
fully occupied initial state and {\it ii)} the survival probability $P_s(t)$ in seed
simulations coincide; hence $\beta=\beta'$. In other realizations of DP
(e.g. site DP), this time reversal symmetry is not exact but only
asymptotically realized.

Applying the same arguments to the present model, the survival probability
$P_s(t)$ in seed simulations should scale in the same way as the density of
active sites at the boundary $\rho_0(t)$ in a process starting with a
\textit{fully occupied lattice} in the bulk. A numerical test, which
approximates such a situation, confirms this conjecture, i.e. one has
$\rho_0(t)\sim t^{-\delta}$ with $\delta \approx 0.15$ for a fully occupied
initial state.

Following the arguments of~\cite{BaratoHinrichsen09} in a related model, this
observation can be used to provide an heuristic explanation for the fact that
$\delta\neq\beta/\nu_\parallel$.

It is known that, if the boundary acts as a sink or perfect trap (e.g.  if
$p=0$), then, in a process starting with a fully occupied lattice, one
observes a growing depletion zone around the boundary whose linear size
$l'(t)$ increases as $l'\sim t^{\alpha_{l}}$, with $\alpha_{l}=1/2$ (see
\cite{rabbits} and the next subsection). Thus, the density of active sites
decays as $t^{-1/2}$. Hence, the influx of particles from the bulk to the
leftmost site may be considered as an effective time-dependent external field
$h(t)\sim t^{-1/2}$. Making the assumption that this field varies so slowly
that the response of the process (i.e. the actual average activity at the
boundary) behaves adiabatically, as if the field was constant, then in a
critical process starting from an initially fully occupied state:
\begin{equation}
  \rho_0(t) \sim t^{-\frac{1}{2\delta_h}} \sim t^{-1/6}.
\end{equation}
Owing to the time reversal property, this quantity should decay as the
survival probability. This chain of heuristic arguments leads to the
conjecture that the survival exponent is given by $ \delta = 1/6$, in
agreement with the numerical estimate $\delta=0.15(2)$.

This unusual value of the exponent $\delta$ is clearly related to the fact
that the present problem is inhomogeneous. The argumentation presented above
does not work for the CP, for example, since there is no special site and,
therefore, a fully occupied lattice cannot be interpreted as a time dependent
field acting on a special site.

\subsection{Density profile}
\label{density}
\begin{figure}
\includegraphics[width=87mm]{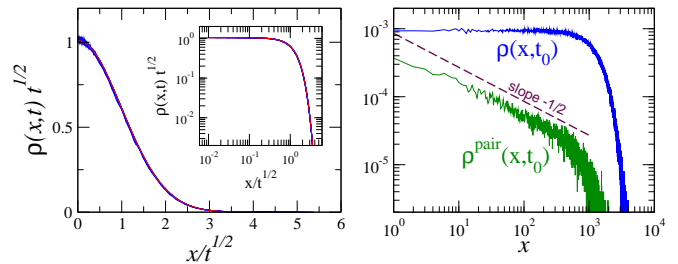}
\vspace{-2mm}
\caption{(Color online) Left: Data collapse of the rescaled profiles of the
  particle density at criticality for $t_0=64,128,\ldots, 8192$ (blue) compared
  to a Gaussian distribution (red). Inset: The same data collapse in a
  double-logarithmic representation. Right: Density of particles (blue) and
  pairs (green) at $t_0=10^6$, showing the presence of correlations which
  decay in space as $x^{-1/2}$, indicating that $\beta/\nu_\perp=1/2$.
  \label{fig:profile}}
\end{figure}
Now, we consider the density profile $\rho(x,t)$ in the bulk, where $x \in
\mathbb{N}$ is the spatial coordinate (distance to the boundary), computed at
the critical point. In the left panel of Fig.~\ref{fig:profile}, we compare
the data collapse of the curves $\rho(x,t)t^{1/2}$ as a function of
$x/t^{1/2}$ with a Gaussian and observe an excellent agreement, indicating
random-walk like behavior with a dynamical exponent $z=2$.  However, in
contrast to a simple random walk, particles are mutually correlated. This is
illustrated in the right panel of Fig.~\ref{fig:profile}, where the connected
correlation function between two nearest neighbors:
\begin{equation}
  \rho^{\rm pair}(x,t)= \langle\rho(x+1,t)\rho(x,t)\rangle- 
  \langle\rho(x+1,t)\rangle\langle\rho(x,t)\rangle
\end{equation}
in a system at the critical point is plotted against time. One observes an
algebraic decay, $x^{-1/2}$, with distance. According to the standard scaling
theory this implies that $\beta/\nu_\perp=1/2$ , confirming that
$z=\nu_\parallel/\nu_\perp=2$.  Moreover, these results are in full agreement
with field theoretical calculations presented in
Ref.~\cite{DeloubriereWijland02} (see section \ref{vanWijland}), which predict
$z=2$ and $\alpha=1/2$.

\section{Mean field approximation}
\label{mean_field}
Here, we  study mean field approximations at different levels. Let us denote by
$\eta_i$ the probability to find a particle at site~$i$; the temporal
evolution within a simple (one-site) mean field approximation is given by:
\begin{eqnarray}
\frac{d\eta_0}{dt}&=& -(1-p)\eta_0+ \frac{1}{2}\eta_1(1-\eta_0),
\label{eqmf1}\\
\frac{d\eta_1}{dt}&=& p\eta_0(1-\eta_1)+
\frac{1}{2}\left(\eta_{2}+\eta_{0}\eta_1-2\eta_{1}\right),
\label{eqmf2}\\
\frac{d\eta_i}{dt}&=& \frac{1}{2}\left(\eta_{i+1}+
\eta_{i-1}-2\eta_{i}\right),~~ \mbox{ for } i=2,3,\hdots\,
\label{eqmf3}
\end{eqnarray}
Note that the equations for the boundary site and its neighbor,
Eq.~(\ref{eqmf1}) and Eq.~(\ref{eqmf2}), include quadratic terms due to the
exclusion constraint, while the equation for sites at the bulk,
Eq.~(\ref{eqmf3}), describes in this approximation a symmetric random walk,
{\it i.e.} it is a diffusion equation. The critical point within simple mean
field theory (where the equation for $\eta_1$ also becomes a diffusion
equation) is $p_c= 1/2$.

Considering a localized initial condition at the boundary,
$\eta_i=\delta_{i,0}$, after a transient time the densities at sites $0$ and
$1$ should, approximately, coincide.  Therefore, from Eq.~(\ref{eqmf1}) with
$\eta_0 \approx \eta_1$, it follows that, at criticality, $\eta_0\sim
t^{-1/2}$.

In the stationary regime, Eq.~(\ref{eqmf1}) leads to $\eta_0\sim (p-1/2)$ for
$p\ge 1/2$. From these results we have:
\begin{equation}
\alpha^{MF}=1/2\,,\qquad \beta^{MF}=1\,.
\end{equation}
To obtain the survival exponent, $\delta$, we follow the arguments of the
preceding section and study the decay of activity from a fully occupied
lattice, $\eta_i=1$ for all $i$.  Integrating Eqs.~(\ref{eqmf1}),
Eq.~(\ref{eqmf2}) and Eq.~(\ref{eqmf3}) numerically with this initial
condition, we obtain an exponent in agreement with
\begin{equation}
\delta^{MF}= 1/4\,.
\end{equation}
A more accurate approximation can be obtained by keeping the correlation
between the first two sites, which is expected to be more relevant than the
correlation between other neighboring sites.  Such a pair-approximation was
used recently in a model where a boundary site also plays a special
role~\cite{Sugden07}. In this approximation, the master equation reads:
\begin{eqnarray}
\frac{d\sigma_{00}}{dt}&=& (1-p)\sigma_{10}+
\frac{1}{2}[\sigma_{01}(1-\eta_2)- \sigma_{00}\eta_2],
\\ \frac{d\sigma_{01}}{dt}&=& (1-p)\sigma_{11}+
\frac{1}{2}[\sigma_{00}\eta_2-\sigma_{01}(2-\eta_2)],\nonumber
\\ \frac{d\sigma_{10}}{dt}&=& -\sigma_{10}+
\frac{1}{2}[\sigma_{01}-\sigma_{10}\eta_2+\sigma_{11}(1-\eta_2)],\nonumber
\\ \frac{d\sigma_{11}}{dt}&=& p\sigma_{10}- (1-p)\sigma_{11}+
\frac{1}{2}[\sigma_{10}\eta_2-\sigma_{11}(1-\eta_2)],\nonumber
\\ \frac{d\eta_2}{dt}&=& \frac{1}{2}\left(\eta_{3}+ \sigma_{11}+ \sigma_{01}
-2\eta_{2}\right),\nonumber \\ \frac{d\eta_i}{dt}&=&
\frac{1}{2}\left(\eta_{i+1}+ \eta_{i-1}-2\eta_{i}\right) \mbox{ for }
i=3,4,\hdots\,\nonumber
\end{eqnarray}
where $\sigma_{s_0s_1}$ is the probability that the occupation numbers of the
first two sites are $s_0$ and $s_1$. Numerical integration of these equations
leads to an improved critical point estimation, $p_c\approx 0.634$, but to the
same mean-field exponents as above.
\section{Related models and field theoretical approaches}
\label{related_models}

\subsection{Bosonic variant}
\label{bosonic}
The model defined above is fermionic in the sense that each site can be
occupied by, at most, one particle. We now consider a bosonic variant without
such a constraint.  This means that diffusion is independent of the
configuration of particles and that particles can be created at the boundary
site without restriction.  More specifically, the update rules are:
\begin{enumerate}
\item[(a)] A particle is chosen randomly.
\item[(b)] If the particle is located at the leftmost site it can:
create another particle at the leftmost site ($s_0=s_0+1$) at rate $\lambda$,
die ($s_0=s_0-1$) at rate $\sigma$, or
diffuse to the next neighbor at rate $D$.
      \item[(c)] If the particle is located in the bulk, it diffuses to the
        right or to the left at equal rates $D$.
\end{enumerate}
The corresponding master equation is:
\begin{eqnarray}
\frac{dP(\{n\},t)}{dt} &=& \lambda\bigl[(n_0-1)P(n_0-1,...,t)-
  n_0P(\{n\},t)\bigr]\nonumber \\
& +&\sigma\bigl[(n_0+1)P(n_0+1,...,t)- n_0P(\{n\},t)\bigr]\nonumber \\
&+& D \Bigl[ \sum_{\langle ij\rangle}P(...,n_i-1,n_j+1,...,t) \\
&+&P(...,n_i+1,n_j-1,...,t)-2P(\{n\},t) \Bigr] \nonumber
\label{eqboson1}
\end{eqnarray} 
where $P(\{n\},t)$ is the probability to find a given configuration $\{n\}=
n_0, n_1, n_2\ldots$ and the sum runs over all nearest neighbors, $j$, of site
$i$ (recall that site $0$ has only one neighbor). Defining the state vector:
\begin{equation}
|\psi(t)\rangle= \sum_{\{n\}}P(\{n\},t)|\{n\}\rangle,
\end{equation}
where $|\{n\}\rangle=\otimes_i|n_i\rangle$ denotes the usual configuration
basis, the master equation can be expressed in the form
\begin{equation}
\frac{d}{dt} |\psi(t)\rangle= -\hat{H}|\psi(t)\rangle\,,
\end{equation}
where $\hat{H}$ is the time evolution operator. Using bosonic creation and
annihilation operators, defined by $\hat{a}_i|n_i\rangle= n_i|n_i-1\rangle$
and $\hat{a}_i^\dagger|n_i\rangle= |n_i+1\rangle$, the master equation
Eq.~(\ref{eqboson1}) can be shown to correspond to the time evolution
operator:
\begin{eqnarray}
\hat{H}&=& D\sum_{\langle
  ij\rangle}(\hat{a}^\dagger_i-\hat{a}^\dagger_j)(\hat{a}_i-\hat{a}_j)
\label{eqbosonH}
\\ &&+\sigma(\hat{a}_0^\dagger-1)\hat{a}_0 +\lambda
\hat{a}_0^\dagger(1-\hat{a}_0^\dagger)\hat{a}_0.\nonumber
\end{eqnarray}
In this formalism, the expectation value of an operator $\hat{B}$ is given by
$\langle\hat{B}\rangle= \langle 1|\hat{B} |\psi(t)\rangle$ where $\langle 1|=
\sum_{\{n\}}\langle \{n\}|$. As is the case for the bosonic contact process
\cite{Baumann05}, the equations for the time evolution of the density of
particles close. From the Heisenberg equation of motion, $\frac{d\hat{B}}{dt}=
[\hat{H},\hat{B}]$ and Eq.~(\ref{eqbosonH}), one obtains:
\begin{eqnarray} 
\frac{d\rho_0}{dt} &=& D(\rho_1-\rho_0)+ \Delta\rho_0 \label{eqboson1b}
\\ \frac{d\rho_i}{dt} &=& D(\rho_{i+1}+\rho_{i-1}-2\rho_i)\qquad
i=1,2,3\ldots\nonumber
\end{eqnarray}
where $\rho_i(t)= \langle a^\dagger_i(t)a_i(t)\rangle= \langle a_i(t)\rangle$
and $\Delta= \lambda-\sigma$.  Alternatively, one could have written a
Langevin equation equivalent to Eq.(\ref{eqbosonH}), and from it, averaging
over the resulting noise, one readily arrives at the same set of equations
Eq.(\ref{eqboson1b}).

From these equations, we can see that the critical point is $\Delta= 0$, where
Eq.(\ref{eqboson1b}) is a diffusion equation.  In the continuum limit,
Eq.~(\ref{eqboson1b}) reads:
\begin{equation}
\frac{\partial\rho(x,t)}{\partial t}= \frac{\partial^2\rho(x,t)} {\partial
  x^2}+ \Delta\delta(x)\rho(x,t)\,
\label{eqboson2}
\end{equation}
where $x$ is the spatial coordinate and, without loss of generality, we have
set $D=1$. We note that in order to take the continuum limit in equation (\ref{eqboson1b}), a site $-1$, with $\rho_{-1}= \rho_{0}$, has to be introduced, so that appropriate boundary conditions are satisfied. The solution of this inhomogeneous diffusion equation is:
\begin{equation}
\rho(x,t)= \int_0^\infty \delta(\zeta)G(x,\zeta,t)d\zeta+\nonumber
\end{equation}
\begin{equation}
\int_0^t\int_0^\infty\Delta\delta(\zeta)\rho(\zeta,\tau)G(x,\zeta,t-\tau),
d\zeta d\tau \,
\label{eqboson3}
\end{equation}
where $G(x,\zeta,t)= (e^{-(x+\zeta)^2/(4t)}+e^{-(x-\zeta)^2/(4t)})/(\sqrt{\pi t})$ is the Green
function and the first term in the right hand side comes from the initial
condition $\rho(x,0)= \delta(x)$. From Eq.~(\ref{eqboson3}) we have
\begin{equation}
\rho_0(t)= \frac{2}{\sqrt{\pi t}}+2\Delta\frac{d^{-1/2}}{dt^{-1/2}}\rho_0(t) \,
\label{eqboson4}
\end{equation}
where $\rho_0(t)= \rho(0,t)$, and the operator $\frac{d^{-1/2}}{dt^{-1/2}}$,
defined by
\begin{equation}
\frac{d^{-1/2}}{dt^{-1/2}}f(t)= \int_0^t\frac{f(\tau)}{\sqrt{\pi (t-\tau)}}d\tau,
\end{equation}
is a half integral operator \cite{Oldham}.  Equation~(\ref{eqboson4}) involves
(owing to the delta function in the interaction term in Eq.~(\ref{eqboson2}))
only the density at the leftmost site. This justifies the mapping of this
model onto an effective one-site non-Markovian process (see next
section). Using some rules for half integration \cite{Oldham} to solve
Eq.~(\ref{eqboson4}), we find:
\begin{equation}   
\rho_0(t)= \frac{2}{\sqrt{\pi t}}+ 4\Delta\exp(4\Delta^2t)\mbox{erf}(-2\Delta\sqrt{t}),
\label{eqboson5}
\end{equation}
where $\mbox{erf(x)}$ is the error function. This implies that, above the
critical point, $\rho_0$ grows exponentially in the long time limit, and does
not reach a stationary value, {\it i.e.} there is a first order transition
and, hence, $\beta= 0$ in this bosonic model.  From equation
Eq.~(\ref{eqboson5}), we deduce $\beta'=1$ and $\nu_\parallel=2$. We have not
been able to calculate the survival-probability exponent exactly, but
numerical simulations suggest $\delta= 1/4$, in agreement with the mean field
exponent.

\subsection{Partially bosonic variant}
\label{partially_bosonic}

Let us now introduce a {\it partially bosonic} variant of the previous model
by retaining the exclusion constraint only at the boundary, but not in the
bulk.  The rules, in this case, are:
\begin{enumerate}
\item[(a)] A particle is randomly chosen.
\item[(b)] If it is at the leftmost site, it can generate a particle at site
  $1$ (provided that $s_1=0$) with probability $p$ or die ($s_0:=0$) with
  probability $1-p$.
\item[(c)] Particles in the bulk diffuse to the right or to the left with the
  same probability, $1/2$.
\end{enumerate}
Numerical simulations show that this variant exhibits the same critical
behavior as the original model, even if the critical point is shifted to $p_c=
0.6973(1)$. This shows that the fermionic constraint is relevant only at the
boundary, where it induces a saturation of the particle density and leads the
transition to become continuous.

\subsection{Models with pair annihilation at the boundary}
\label{vanWijland}
In the models discussed so far, particles at the boundary either create an
offspring or die spontaneously at some rate. Instead, a very similar model was
introduced in Ref.~\cite{DeloubriereWijland02}, for which particles at the
boundary annihilate only in \textit{pairs}.  In its fermionic variant,
particles at sites $0$ and $1$ annihilate with each other (provided that both
sites are occupied) at some rate, while isolated particles at the boundary
cannot disappear:
\begin{eqnarray*}
\mbox{present models:} & A\to2A \,,\quad A\to\emptyset\,, \\ \mbox{models of
  Ref.~\cite{DeloubriereWijland02}:} & A\to2A \,,\quad 2A\to\emptyset\,.
\end{eqnarray*}
Analogously, one can define a bosonic version, in which two particles at the
boundary can annihilate. In the following discussion we consider these two
variants in $d$ spatial dimensions where, as is the case $d=1$, only a single
site has ``special" dynamics.
 
A detailed field theoretical analysis of these pair-annihilating models was
presented in \cite{DeloubriereWijland02}. In the bosonic case, proceeding as
above (see Eq.(\ref{eqbosonH})) one obtains the following time evolution
operator:
\begin{eqnarray}
\hat{H}&=& D\sum_{\langle ij\rangle}(\hat{a}^\dagger_i-\hat{a}^\dagger_j)
(\hat{a}_i-\hat{a}_j) \nonumber
\\ &&+\sigma[(\hat{a}_0^\dagger)^2-1]\hat{a}_0^2 +\lambda
\hat{a}_0^\dagger(1-\hat{a}_0^\dagger)\hat{a}_0.
\label{H2}
\end{eqnarray}  
which, after eliminating higher order terms and taking the continuum limit, is
equivalent to a Langevin equation identical to the one for DP except for the
fact that
all terms, except for the Laplacian, are multiplied by a $\delta$ function at
the boundary; {\it i.e.} the non-diffusive part of the dynamics operates only
at the boundary.  An $\epsilon$-expansion analysis of Eq.(\ref{H2}) (see
\cite{DeloubriereWijland02}) leads to $\alpha=1/2$ and $z=2$ as exact results
in all orders of perturbation theory, and to $\beta=1-3(4-3d)/8$, up to first
order in $\epsilon =4/3-d$ around the critical dimension $d_c=4/3$.  Also, it
was shown that the time reversal symmetry is preserved.

\begin{figure}[t]
\begin{center}
  \includegraphics[height=5cm]{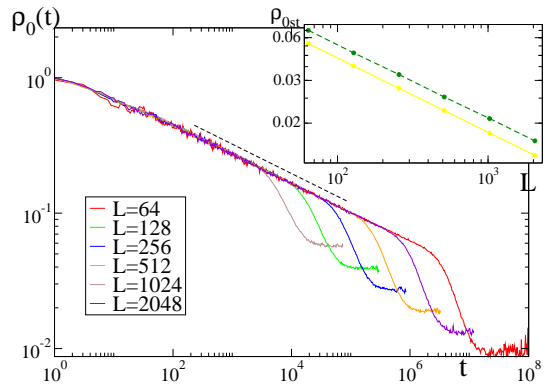}
\end{center}
\caption{\footnotesize{(Color online) Temporal behavior of $\rho_{0}$ for the bosonic
    pair-annihilating model, starting from a homogeneous initial condition for
    different system sizes (from $L=64$ to $L=2048$).  The exponent
    $\beta/\nu_{\perp}$ can be measured from the scaling of the different
    saturation values as a function of system size (see inset; yellow
    line). Also, in the inset (dashed green line), we show the scaling of saturation
    values for the fermionic version of the same model, showing the same type
    of scaling. }}
\label{FSS}
\end{figure}

We have verified all these predictions in computer simulations of the bosonic
annihilation model. For instance, from the time decay of $\rho_0(t)$, as shown
in Fig.~\ref{FSS}, we determine $\delta = 0.21(3)$, while from a finite size
scaling analysis of the saturation values of the order parameter at
criticality we measure $\beta/\nu_{\perp}=0.51(2)$ (see Fig.~\ref{FSS}), in
reasonable agreement with the expected results, $\delta= 1/6$ and
$\beta/\nu_{\perp}=1/2$, respectively. Moreover, from spreading simulations
(not shown) we estimate $\alpha \approx 1/2 $ and $z \approx 2$. All the
exponents are in agreement with the ones presented in the previous section for
single particle annihilation models.

Actually, a simple argument explains why the model of section
\ref{simulations} and the pair-annihilation model share the same
critical behavior. This is plausible because the chain reaction $A \to
2A \to \emptyset$ in the model with pair annihilation generates
effectively the reaction $A\to\emptyset$ of the model considered with
CP-like dynamics.

Hence, the field theoretical predictions discussed above
\cite{DeloubriereWijland02,BaratoHinrichsen08} apply also to the CP-like
model. In $d=1$, the one-loop prediction $\beta= 5/8= 0.625$
\cite{DeloubriereWijland02}, is not far from the exponent measured in section
\ref{simulations}, $\beta=0.68(5)$.

On the other hand, the fermionic version of the pair-annihilating model has
been conjectured to yield in a different universality class, and a prediction
for its critical exponents is made in \cite{DeloubriereWijland02} (for
instance, $\beta=1$).  Our numerical simulations disprove such a claim; all
the measured critical exponents for the fermionic variant of the
pair-annihilation model are numerically indistinguishable from their bosonic
counterparts (see Fig.~\ref{FSS}).

In summary, all the defined models, either with single particle annihilation
or with pair-annihilation, fermionic or bosonic, exhibit a boundary induced
phase transitions and, except for one of them, they all are continuous and
share the same critical behavior. The exception to this rule is the CP-like
model without a fermionic constraint at the boundary, which lacks of a
saturation mechanism in the active phase, leading to unbounded growth of 
particle density at the leftmost site above the critical point and to a discontinuous transition.
\section{Relation to a $(0+1)$-dimensional non-Markovian process}
\label{non-Markovian}

In Ref.~\cite{DeloubriereWijland02}, by integrating out the fields related to
diffusion in the bulk from the corresponding action, it was shown that the
class of boundary-induced phase transitions into an absorbing state considered
here can be related to a non-Markovian single site process. The properties of
such a spreading process on a time line has been studied in further detail in
Ref.~\cite{BaratoHinrichsen09}.

On an heuristic basis, the relation can be explained as follows: consider the
CP-like model only from the perspective of the leftmost site. A particle at
the origin may die or create a new particle that will go for a random walk
coming back to the origin after a time $\tau$. What happens during this random
walk is irrelevant from the perspective of the leftmost site; the only
relevant aspect is the time needed for a created particle to come back to the
boundary.  Once it returns it may die or create new offsprings which, on their
turn, will undergo random walks in the bulk.

Our simulations above show that the fermionic constraint is irrelevant in the
bulk. Therefore, we can consider without lost of generality the bulk-bosonic
version in which there is no effective interaction among diffusing
particles. In this case, the probability distribution of the returning time to
the origin has the well-known asymptotic form~\cite{Redner01}:
\begin{equation}
\label{WaitingTimeDistribution}
P(\tau)\sim \tau^{-3/2}\,.
\end{equation}
Taking all these elements into account we define the following
non-Markovian model on a single site~\cite{BaratoHinrichsen08}:
\begin{enumerate}
\item[(a)] Set initially $s(t):= \delta_{t,0}$ for all times, $t$.
\item[(b)] Select the lowest $t$ for which $s(t)=1$.
\item[(c)] With probability $\mu$, generate a waiting time $\tau$ according to
  the distribution~Eq.~(\ref{WaitingTimeDistribution}), truncate it to an
  integer, and set $s(t+\tau):=1$; otherwise (with probability $1-\mu$) set
  $s(t):=0$.
\item[(d)] Go back to (b).
\end{enumerate}
The process runs until the system enters the absorbing state ($s(t')=0$ for
all $t'>t$) or a predetermined maximum time is exceeded.
\begin{figure}
\includegraphics[width=87mm]{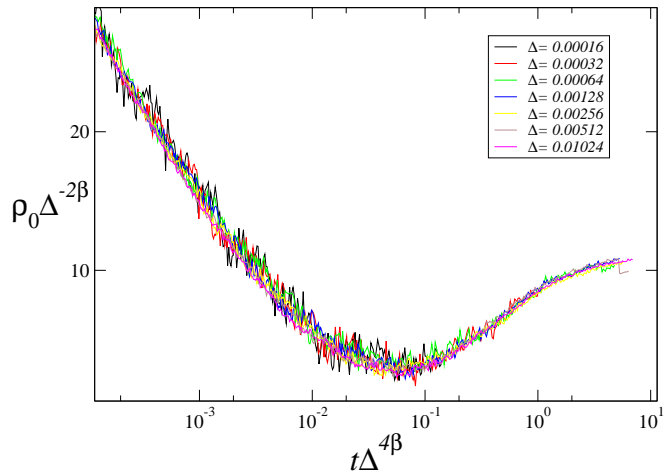}
\vspace{-2mm}
\caption{(Color online) Off-critical data collapse with the one-site model:
  $\langle s(t)\rangle\Delta^{-2\beta}$ as a function of $t\Delta^{4\beta}$
  for different values of $\Delta$, with $\beta= 0.71(2)$.
  \label{fig:datacollapse}}
\end{figure}
The density of particles at the leftmost site of the original model is related
to $\langle s(t)\rangle$ in the single-site model, the survival probability at
time $t$ is given by the fraction of runs surviving at least up to $t$, and
the initial condition $s(t):= \delta_{t,0}$ corresponds to start with a single
particle at the boundary in the full model.  Critical exponents can be defined
as in the original model. However, the simulation results for the single-site
non-Markovian model are more reliable because it is possible to perform much
longer runs and, in the case of off-critical simulations, one can work with
smaller values of $\Delta$. With time-dependent simulations at the critical
point $\mu_c=0.574262(2)$, we obtained $\alpha=0.500(5)$ and
$\delta=0.165(3)$, in good agreement with the conjectured values $\alpha= 1/2$
and $\delta= 1/6$.  As an example, we show the results of supercritical
simulations in Fig.~\ref{fig:datacollapse}, where we obtained a convincing
data collapse by plotting $\langle s(t)\rangle\Delta^{-2\beta}$ as a function
of $t\Delta^{4\beta}$ for different values of $\Delta$ with $\beta=
0.71(2)$. The latter estimate is in agreement with $\beta= 0.68(5)$, coming
from the original model.

As shown in previous studies (see e.g.~\cite{Hinrichsen07} and references
therein), a non-Markovian time evolution with algebraically distributed
waiting times $P(\tau) \sim \tau^{-1-\kappa}$ is generated by so-called
fractional derivatives $\partial_t^\kappa$ which are defined by:
\begin{equation}
\label{eq:IntegralTime}
\partial_t^\kappa \, \rho(t)  \;=\;  \frac{1}{\mathcal{N}_\parallel(\kappa)}
\int_0^{\infty} {\rm d}t' \, {t'}^{-1-\kappa} [\rho(t)-\rho(t-t')]\,,
\end{equation}
where $\kappa\in[0,1]$ and $\mathcal{N}_\parallel(\kappa)=-\Gamma(-\kappa)$ is
a normalization constant. Hence, we expect this model to be described by a
DP-like $0$-dimensional Langevin equation with a half-time derivative, instead
of the usual one, to account for the non-Markovian character of the model:
\begin{equation}
\label{Langevin}
\partial_t^{1/2} \rho(t) = a \rho(t) - \rho(t)^2 + \xi(t)\,
\end{equation}
where $a$ is proportional to the distance from criticality and $\xi$ is a
multiplicative noise with correlations $\langle \xi(t)
\xi(t')\rangle=\rho(t)\delta(t-t')$. This equation can be obtained from the
effective action that arises when the fields related to diffusion in the bulk
are integrated out, and the relation of the order of the fractional derivative
in a generalized one-site model with the dimension in the full model is
$\kappa= (2-d)/2$ \cite{DeloubriereWijland02}. An analysis of this one-site
model with general $\kappa$ and a comparison with the results coming from
field theory is presented in \cite{BaratoHinrichsen09}.
\section{Conclusion}
\label{conclusions}

We have studied boundary-induced phase transitions into an absorbing
state in one-dimensional systems with creation/annihilation dynamics
at the boundary and simple diffusive dynamics in the bulk.  The
non-trivial dynamics at the boundary induces a phase transition in the
bulk. We have analyzed such a transition for different though similar
models, including different ingredients: either single-particle
annihilation or pairwise annihilation, fermionic constraint or lack of
it, etc.

A particular bosonic version can be exactly solved; owing to the lack of any
saturation mechanism, the density of particles grows unboundedly in the active
phase, leading to a discontinuous transition with trivial critical exponents.

The rest of the analyzed models exhibit a continuous transition and define a
unique universality class. At the bulk, the dynamics is governed by
random-walks, entailing the exponent values $z=2$ and $\alpha=1/2$. On the
other hand, some critical exponents take non-trivial values:
{\it i)} the survival probability from a localized seed at
the boundary exponent, which from an heuristic argument supported by
simulations results, turns out to be $\delta=1/6$, as well as {\it ii)} the
order parameter exponent, $\beta= 0.71(2)$.  The remaining exponents can be
obtained from these ones using scaling relations.

Finally, it has been shown that the class of boundary induced phase
transitions studied here can be related to a single-site non-Markovian
process.  This process is particularly suitable for numerical
simulations and it is also of conceptual interest in the sense that it
shows that nonequilibrium phase transitions can occur even in $0+1$
dimensions by choosing an adequate non-Markovian dynamics. It is also
convenient for the comparison of the results obtained form the
$\epsilon$-expansion and simulations
\cite{BaratoHinrichsen09,DeloubriereWijland02}.

The models studied here possibly constitute the simplest universality class of
nonequilibrium phase transition into an absorbing state, in the sense that the transition occurs 
because of the special dynamics of just one site and, in contrast to DP, some critical exponents
can be obtained exactly from the field theory.

\begin{acknowledgments}
  We thank X. Durang and M. Henkel for helpful discussions. 
  Financial support by the Deutsche Forschungsgemeinschaft (HI
  744/3-1), by the Spanish MEyC-FEDER, project FIS2005-00791, and from
  Junta de Andaluc{\'\i}a as group FQM-165 is gratefully acknowledged.
\end{acknowledgments}


\end{document}